\documentclass[superscriptaddress,twocolumn,showpacs,prb,floatfix]{revtex4}
\usepackage{epsfig}

\begin{document}

\title{Dynamical scaling in Ising and vector spin glasses}

\author{Helmut G.~Katzgraber}
\affiliation{Theoretische Physik, ETH H\"onggerberg,
CH-8093 Z\"urich, Switzerland}

\author{I.~A.~Campbell}
\affiliation{Laboratoire des Verres, Universit\'e Montpellier II,
34095 Montpellier, France}

\date{\today}

\begin{abstract}
We have studied numerically the dynamics of spin glasses with Ising and $XY$ 
symmetry (gauge glass) in space dimensions 2, 3, and 4. The nonequilibrium 
spin-glass susceptibility $\chi_{\rm ne}(t_{\rm w},T)$ and the 
nonequilibrium energy per spin $e_{\rm ne}(t_{\rm w},T)$ of samples of 
large size $L_{\rm b}$ are measured as a function of anneal time $t_{\rm 
w}$ after a quench to temperatures $T$. The two observables are compared to 
the equilibrium spin-glass susceptibility $\chi_{\rm eq}(L,T)$ and the 
equilibrium energy $e_{\rm eq}(L,T)$, respectively, measured as functions 
of temperature $T$ and system size $L$ for a range of system sizes. 
For any time and temperature a nonequilibrium time-dependent length 
scale $L^{*}(t_{\rm w},T)$ can be defined by writing $\chi_{\rm 
ne}(t_{\rm w},T) = \chi_{\rm eq}(L^{*},T) $ (or the equivalent expression 
for the energy). Our analysis shows that for all systems studied, an 
``effective dynamical critical exponent'' parametrization $L^{*}(t_{\rm 
w},T)= A(T)t^{1/z(T)}$ fits the data well at each temperature within the 
whole temperature range studied, which extends from well above the critical 
temperature $T_{\rm c}$ to near $T=0$ for dimension 2 or to well below 
$T_{\rm c}$ for the other space dimensions studied. In addition, the data
suggest that the dynamical exponent $z$ varies smoothly when crossing
the transition temperature.
\end{abstract}

\pacs{75.50.Lk, 75.40.Mg, 05.50.+q}
\maketitle

\section{Introduction}
\label{introduction}

The dynamics of laboratory spin glasses manifests a number of fascinating 
phenomena, linked to the fact that below the glass temperature $T_{\rm c}$ the 
systems never achieve true thermodynamic equilibrium.\cite{binder:86,young:98}
It has gradually become clear that a slow increase of the coherence 
length (also known as ``dynamic correlation length'') with time plays an 
important role in the memory and rejuvenation 
effects seen experimentally in spin glasses under various cooling and heating 
protocols.\cite{dupuis:01,bouchaud:02,jonsson:02,bert:04,jimenez:04,berthier:05}

Numerical work can also bring light to bear on the question. 
Finite-size scaling theory states that the dependence of the equilibrium 
spin-glass susceptibility $\chi_{\rm eq}$ on the system size $L$ at 
the critical temperature $T_{\rm c}$ is given by
\begin{equation}
\chi_{\rm eq}(L) \sim L^{2-\eta} \;\;\;\;\;\;\;\;\; (T = T_{\rm c}),
\end{equation}
where $\eta$ is the ``anomalous dimension'' static scaling exponent for the 
correlation function of the 
system. In addition, at $T_{\rm c}$, 
the equilibrium autocorrelation relaxation time 
increases with sample size $L$ as
\begin{equation}
\tau(L) \sim L^{z_{\rm c}} \; , 
\end{equation}
where $z_{\rm c}$ is the dynamical critical exponent, as conventionally 
defined for the standard single-spin Glauber update dynamics. 
Huse\cite{huse:89} remarked 
that the critical anneal-time dependence of the nonequilibrium spin-glass 
susceptibility for large samples after a quench to $T_{\rm c}$ is
\begin{equation} 
\chi_{\rm ne}(t_{\rm w}) \sim t_{\rm w}^{(2-\eta)/z_{\rm c}} \; ,
\label{eq:chi_ne}
\end{equation}
where $t_{\rm w}$ is the time after quench and $z_{\rm c}$ is 
again the equilibrium 
dynamical critical exponent. Equation (\ref{eq:chi_ne})
is strictly equivalent to the 
definition of an effective time-dependent length scale via
\begin{equation}
L^{*}(t_{\rm w}) = A_{\rm c}t_{\rm w}^{1/z_{\rm c}}\; ,
\label{crit_scale}
\end{equation}
with $A_{\rm c}$ an appropriate prefactor. The fact that the nonequilibrium 
scaling should depend only on the equilibrium dynamical critical
exponent $z_{\rm c}$ is 
important and nontrivial. In the first case the system is quenched from 
infinite temperature; thus its effective temperature is changing with time 
all through the annealing process. In the second case $z_{\rm c}$ 
represents the 
dynamical scaling for thermal fluctuations within the set of 
configurations at thermodynamic equilibrium. A rigorous theoretical 
justification for the nonequilibrium approach exactly at criticality has 
been given by Jannsen {\em et al}.\cite{jannsen:89} Nonequilibrium dynamics 
has been studied numerically in considerable detail for many regular 
magnetic systems,\cite{comment:convention} principally because such data 
can give accurate information on the critical behavior (see, for instance, 
Ref.~\onlinecite{zheng:99}).

Spin-glass critical nonequilibrium behavior has already been studied 
numerically in a number of Ising spin glasses (ISG's) and the gauge glass
(GG).\cite{huse:89,blundell:92,mari:02,bernardi:97,katzgraber:04f} 
Relaxation becomes very slow below $T_{\rm c}$ in glassy systems; 
the phrase ``time is length'' has been coined for the link between 
coherence length $\ell(t_{\rm w})$ and anneal time 
$t_{\rm w}$,\cite{berthier:02} and the fact has 
been underlined that the physically relevant length scales involved in the 
dynamics of spin glasses below $T_{\rm c}$ are short, even for experimental 
times which are always very long compared to microscopic time scales. This 
is because the values of $z_{\rm c}$ in spin glasses are intrinsically 
high, as the mean field $z_{\rm c}$ at the ISG upper critical 
dimension $d=6$ is already 
equal to $4$, and $z_{\rm c}(d)$ increases to yet higher values at lower 
dimensions.

In numerical simulations of ISG's below $T_{\rm c}$ the time dependence of the 
dynamical correlation length $\ell(t_{\rm w})$ has been 
estimated\cite{parisi:96,kisker:96,berthier:02,yoshino:02} by measuring the 
time-dependent correlation function explicitly and parametrizing using an 
appropriate assumption for the form of the function. Early correlation 
length data were analyzed using the phenomenological assumption that a 
dynamical scaling relationship of the critical functional form, 
Eq.~(\ref{crit_scale}), continues to hold for $\ell(t_{\rm w})$ at temperatures 
below $T_{\rm c}$, with a temperature-dependent effective dynamical exponent 
$z(T)$.\cite{kisker:96,parisi:96}
It has been suggested that $z(T)\propto 1/T$.\cite{parisi:96}

Alternatively a ``dynamic droplet scaling'' has been proposed, where below 
$T_{\rm c}$ an excitation barrier increases algebraically with correlation 
length.\cite{fisher:88b,fisher:88} For ISG's in three and four dimensions
the consequences 
of an analysis based on this approach have been discussed in 
detail in Refs.~\onlinecite{berthier:02}, \onlinecite{kisker:96},
and \onlinecite{yoshino:02}. 
This form of parametrization has 
been widely employed in analyses of experimental 
data\cite{mattsson:95,dupuis:01,bouchaud:02a,jonsson:02a,bert:04} although it 
should be noted that length scales are never measured directly in 
experiments.

In the paramagnetic state ($T > T_{\rm c}$), 
relaxation is fast compared with the time 
scales of most experiments; therefore measurements of relaxation rates are very 
difficult. No numerical studies of dynamical length scales seem to have been 
undertaken either in the regime above the spin-glass freezing temperature 
in systems having a finite $T_{\rm c}$, 
except for the two-dimensional (2D) Edwards-Anderson 
(EA) ISG for which $T_{\rm c}=0$, where careful studies have been 
undertaken.\cite{kisker:96,rieger:04}

In this work we present results from simulations of the EA ISG with 
Gaussian-distributed interactions, as well as
the GG in dimensions 2, 3, and 4. We choose these models as they are
paradigmatic representatives of spin glasses with Ising as well as with
vector spin symmetry.
We define a temperature-dependent length scale $L^{*}(t_{\rm w},T)$ and 
find that the equation having the same functional form as
Eq.~(\ref{crit_scale}) valid at criticality, 
\begin{equation}
B
L^{*}(t_{\rm w},T) = A(T)t_{\rm w}^{1/z(T)} \; ,
\label{eff_exponent}
\end{equation}
with a temperature-dependent effective exponent $z(T)$ and a weakly 
temperature-dependent prefactor $A(T)$, gives an excellent parametrization 
of the data in each system, not only at temperatures below $T_{\rm c}$ 
(confirming the conclusions of Refs.~\onlinecite{parisi:96} and
\onlinecite{kisker:96}), but also above $T_{\rm c}$. 
In addition, we find a disagreement with ``droplet dynamic 
scaling''\cite{berthier:02,yoshino:02} for the Ising spin glasses, 
which becomes apparent for temperatures below $\sim T_{\rm c}/2$. 
For temperatures above $T_{\rm c}$ neither 
the droplet approach nor the replica symmetry breaking (RSB) approach 
appears to give any predictions as to a dynamic scaling. For both 
pictures, ``barriers'' in the energy landscape
disappear above $T_{\rm c}$, therefore relaxation becoming trivial.

In Sec.~\ref{definition_length} we define and discuss
a time-dependent length scale which
will be needed to rescale the data. In Secs.~\ref{sec:gg} and \ref{sec:isg}
we present data for the two-, three-, and four-dimensional gauge glass and
Ising spin glass, respectively. After a brief summary
(Sec.~\ref{sec:summary}), droplet dynamic scaling is discussed in 
Sec.~\ref{sec:droplets}, followed by concluding remarks in 
Sec.~\ref{sec:concl}.

\section{Spatiotemporal scaling}
\label{definition_length}

The relationship between time and length is defined and discussed below
using the internal energy and the spin-glass susceptibility. In
general, the internal energy for a spin Hamiltonian is given by
\begin{equation}
e = \frac{1}{N} [\langle {\mathcal H}\rangle]_{\rm av} \; .
\label{eq:mean_energy}
\end{equation}
Here $N = L^d$ represents the number of spins described by a Hamiltonian 
${\mathcal H}$ on a hypercubic lattice of linear size $L$ 
and $\langle \cdots \rangle$ represents a thermal average,
whereas $[\cdots]_{\rm av}$ corresponds to a disorder average.

The spin-glass susceptibility (related to the nonlinear
susceptibility measured experimentally) can be expressed as
\begin{equation}
\chi = N [\langle |q|^2\rangle]_{\rm av} \; .
\label{eq:mean_susc}
\end{equation}
Here $q$ represents the Edwards-Anderson order parameter which in the case of
the gauge glass is given by
\begin{equation}
q = \frac{1}{N} \sum_{j=1}^N \exp[i(\phi_j^{\alpha} - \phi_j^{\beta})]\; ,
\label{eq:q_gg}
\end{equation}
with $\phi_j$ representing the phases of the $XY$ spins, and $\alpha$ and $\beta$
are two copies of the system with the same disorder. For the Ising spin
glass $q$ is given by
\begin{equation}
q = {1 \over N} \sum_{j=1}^N S_j^{\alpha} S_j^{\beta} ,
\label{q}
\end{equation}
where $S_j = \pm 1$ represent Ising spins.

\subsection{Definition of a length scale}
\label{definition_length.a}

For any fixed $T$, as the system size $L$ is increased or the anneal time 
$t_{\rm w}$ lengthed, the SG equilibrium and nonequilibrium susceptibilities 
$\chi_{\rm eq}(L,T)$ and $\chi_{\rm ne}(t_{\rm w},T)$ 
grow while the internal 
equilibrium and nonequilibrium energies per spin, $e_{\rm eq}(L,T)$ and 
$e_{\rm ne}(t_{\rm w},T)$, respectively, 
drop towards the infinite-size equilibrium value 
$e_{\rm eq}(\infty,T)$. If $T$ is above the ordering temperature, 
$\chi_{\rm eq}(L,T)$ and $\chi_{\rm ne}(t_{\rm w},T)$ will 
both saturate at a 
temperature-dependent limiting value $\chi_{\rm eq}(\infty,T)$; 
$e_{\rm eq}(L,T)$ and 
$e_{\rm ne}(t_{\rm w},T)$ will always saturate at $e_{\infty}(T)$ for all $T$.

Quite generally, if at any temperature $T$ the equilibrium SG 
susceptibility as a function of sample size $L$ is $\chi_{\rm eq}(L,T)$ and 
the nonequilibrium SG susceptibility after an anneal time $t_{\rm w}$ following 
a quench to temperature $T$ is $\chi_{\rm ne}(t_{\rm w},T)$ 
for a large sample of 
size $L_{\rm b}$, then an infinite-sample-size time-dependent length scale 
$L^{*}(t_{\rm w})$ can be rigorously defined by writing
\begin{equation}
\chi_{\rm eq}[L^{*}(t_{\rm w}),T] = \chi_{\rm ne}(t_{\rm w},T) \; ,
\label{chi_chi}
\end{equation}
as long as $L^{*}(t_{\rm w},T) \ll L_{\rm b}$. 
This length scale definition applies 
to any temperature and any anneal time, and the length scale 
$L^{*}(t_{\rm w},T)$ can be estimated numerically to high precision if data of 
sufficient statistical accuracy are available. A practical limitation to 
precision under some conditions is the need to intrapolate between the 
$\chi_{\rm eq}(L,T)$ sequence of values for integer $L$ to provide a 
continuous function with which to compare the time-dependent data. In the 
paramagnetic regime, as $\chi_{\rm eq}(L,T)$ and $\chi_{\rm ne}(t_{\rm w},T)$ 
both saturate at the equilibrium infinite-size 
$\chi_{\rm eq}(\infty,T)$ 
value, $L^{*}(t_{\rm w},T)$ can only be determined up to some 
temperature-dependent finite time.

An analogous equation can be derived for the energy per spin,
\begin{equation}
e_{\rm eq}[L^{*}(t_{\rm w}),T] = e_{\rm ne}(t_{\rm w},T)\; ,
\label{e_e}
\end{equation}
providing an independent estimate for $L^{*}(t_{\rm w},T)$. 
Later, we show that similar estimates of $L^{*}(t_{\rm w},T)$ 
are obtained from a completely independent analysis of the data for the 
susceptibility $\chi$ and energy $e$.

Note that $L^{*}(t_{\rm w},T)$ is not the time-dependent coherence 
length $\ell(t_{\rm w},T)$,\cite{berthier:02} 
but it is closely related to it. During the anneal each 
individual spin is surrounded by a growing cohort of spins in equilibrium 
correlation with it at the temperature $T$, up to a time-dependent cutoff 
length. 
This correlated volume can be considered equivalent to a box of 
linear size $L^{*}(t_{\rm w},T)$. Because $\ell(t_{\rm w},T)$ 
is defined as a correlation 
length and $L^{*}(t_{\rm w},T)$ from the box size, we should expect 
$L^{*}(t_{\rm w},T)\simeq 2\ell(t_{\rm w},T)$. Here, $\ell(t_{\rm w},T)$ 
has been estimated from 
directly measured correlation functions, although a fully rigorous 
definition is not easy to give because of nontrivial  
prefactors.\cite{kisker:96,parisi:96,berthier:02,yoshino:02}
For the 3D EA ISG where 
the comparison can be made directly, the equivalence between 
$L^{*}(t_{\rm w},T)$ and $2\ell(t_{\rm w},T)$ is indeed found to be 
correct. It should be borne in mind that because of the disorder, the length 
values represent averages over different samples and the effective length 
scale $L^{*}(t_{\rm w},T)$ may be slightly different depending over which 
observable the mean is being taken.

\subsection{Comparison between effective length scales of different observables}
\label{comp}

There is a straightforward way to test the assumption that the observables 
$\chi_{\rm ne}(t_{\rm w},T)$ and $e_{\rm ne}(t_{\rm w},T)$ 
are controlled by a single length 
scale. At each $T$, the equilibrium energy $e_{\rm eq}(L,T)$ 
can be written as a 
function of the equilibrium susceptibility $\chi_{\rm eq}(L,T)$ with 
$L$ as an implicit parameter: 
$e_{\rm eq}(L,T) = {f}_e[\chi_{\rm eq}(L,T)]$.  
Suppose that Eqs.~(\ref{chi_chi}) and 
~(\ref{e_e}) hold, with one and the same $L^{*}(t_{\rm w},T)$ for both 
observables. Then the nonequilibrium energies $e_{\rm ne}(t_{\rm w},T)$ 
are given by the identical 
function ${f}_e[\chi_{\rm ne}(t_{\rm w},T)]$ of the nonequilibrium
susceptibility $\chi_{\rm ne}(t_{\rm w},T)$, with 
$t_{\rm w}$ 
an implicit parameter (see Fig.~\ref{fig:D1}). 
A plot of nonequilibrium 
$[e_{\rm ne}(t_{\rm w},T)-e_{\rm eq}(\infty,T)]$ 
against $\chi_{\rm ne}(t_{\rm w},T)$ data is 
superimposed on a plot of $[e_{\rm eq}(L,T) - 
e_{\rm eq}(\infty,T)]$ 
against $\chi_{\rm eq}(L,T)$ data [by definition the equilibrium and 
nonequilibrium energy measurements tend to the identical 
$e_{\rm eq}(\infty,T)$ which we estimate by extrapolation]. These figures are 
simple displays of raw data, and no fitting procedure whatsoever is 
involved; at this stage no assumption is made as to the functional form of 
$L^{*}(t_{\rm w},T)$.

In what follows we present data for the gauge glass, as well as the Ising spin
glass.

\section{Gauge Glass}
\label{sec:gg}

The gauge glass is a canonical vector spin glass (see, for instance, 
Refs.~\onlinecite{kosterlitz:98,olson:00,akino:02}) where $XY$ spins on a 
(hyper)cubic 
lattice in $d$ dimensions of size $L$ interact through the Hamiltonian
\begin{equation}
{\cal H} = -J \sum_{\langle i, j\rangle} \cos(\phi_i - \phi_j - A_{ij}) \; ,
\label{hamiltonian_gg}
\end{equation}
the sum ranging over nearest neighbors. The angles $\phi_i$ represent the 
orientations of the $XY$ spins, and the $A_{ij}$ are quenched random 
variables uniformly distributed between $[0,2\pi]$ with the constraint 
that $A_{ij} = -A_{ji}$ (here $J = 1$). Periodic 
boundary conditions are applied. The GG ordering temperatures have been 
shown to be $T_{\rm c} = 0$, $0.46(1)$, and $0.89(1)$ in dimensions 
$2$, $3$, and $4$, 
respectively.\cite{simkin:96,granato:98,olson:00,katzgraber:03a,katzgraber:04f}

For $XY$ spin systems there is a choice to be made in the allowed 
single-spin acceptance angle for individual updating steps. To optimize 
the updating procedure at low temperatures, the limiting angle is 
often chosen to be less than 
$2 \pi$ for an $XY$ spin\cite{katzgraber:01a} and linearly dependent on 
temperature. The numerical prefactor for the temperature-dependent window 
is chosen so that the acceptance ratios for the local Monte Carlo updates 
is $\sim 0.4$. As far as the final equilibrium parameters are concerned, 
this choice plays no role. However, for the nonequilibrium simulations it 
is essential to use the full $2\pi$ acceptance angle window to obtain 
physically significant results as otherwise the limited angle introduces
an artificial temperature variation in the relaxation.

\subsection{Two dimensions}
\label{sec:2dgg}

The GG in space dimension 2 has a zero-temperature ordering 
transition.\cite{fisher:91,gingras:92,reger:93,simkin:96,granato:98,akino:02,katzgraber:02a,katzgraber:03a} Therefore
all the measurements discussed in this section 
necessarily refer to the paramagnetic state.  Dimension $d=2$ presents the 
advantage that systems can be equilibrated up to large $L$, so in at least 
part of the temperature range comparisons can be made between 
$\chi_{\rm ne}(t_{\rm w},T)$ and $\chi_{\rm eq}(L,T)$ 
over a wide range of system sizes.\cite{comment:figs}
Details of the simulations are presented in the Appendix, Table
\ref{tab_gg_eq_2} for equilibrium and
Table \ref{tab_gg_neq} for nonequilibrium measurements, respectively. 

\begin{figure}
\includegraphics[width=8.0cm]{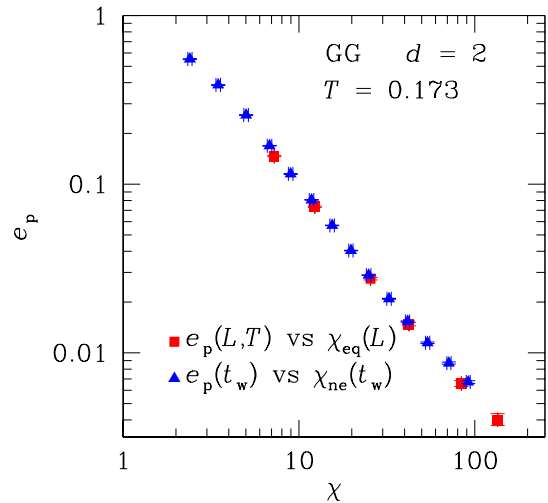}
\vspace*{-1.0cm}
\caption{(Color online)
Energy difference between the energy per 
spin for a large sample ($L=128$) at time $t_{\rm w}$ after quench and the 
infinite time energy, Eq.~(\ref{eq:ep1}), plotted 
against the large-sample spin-glass susceptibility after quench 
$\chi_{\rm ne}(t_{\rm w})$ (triangles)
for the two-dimensional GG at $T = 0.173$.
On the same plot the equilibrium energy difference 
for size $L$ [Eq.~(\ref{eq:ep2})] is plotted against the equilibrium 
susceptibility $\chi_{\rm eq}(L)$ for the same size (squares). 
These are raw data, 
and no fitting is involved in these plots.}
\label{fig:D1}
\end{figure}

\begin{figure}
\includegraphics[width=8.0cm]{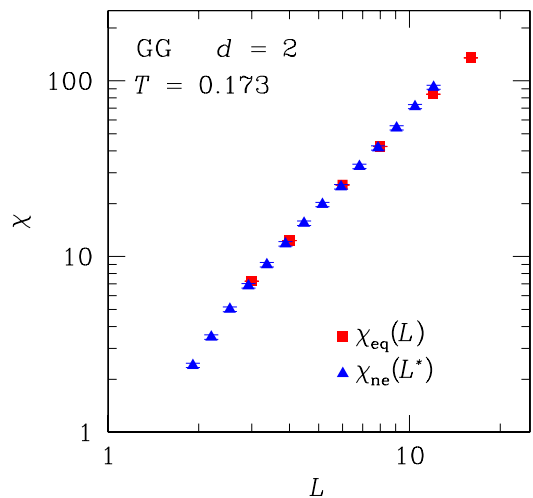}
\vspace*{-1.0cm}
\caption{(Color online)
Finite-size 
equilibrium susceptibilities $\chi_{\rm eq}(L)$ vs 
$L$ (squares) for the two-dimensional GG at $T=0.173$. 
In addition, the susceptibilities $\chi_{\rm ne}(t_{\rm w})$ at time 
$t_{\rm w}$ after a
quench are plotted against 
$L^{*}(t_{\rm w}) = A(t_{\rm w})t_{\rm w}^{1/z(T)}$ 
with the parameters $A$ and $z$, both temperature dependent, 
adjusted 
for optimal overall overlap with the equilibrium data points 
(triangles). In this figure $L^*(t_{\rm w}) = 1.66 t_{\rm w}^{1/4.9}$.}
\label{fig:D2}
\end{figure}

\begin{figure}
\includegraphics[width=8.0cm]{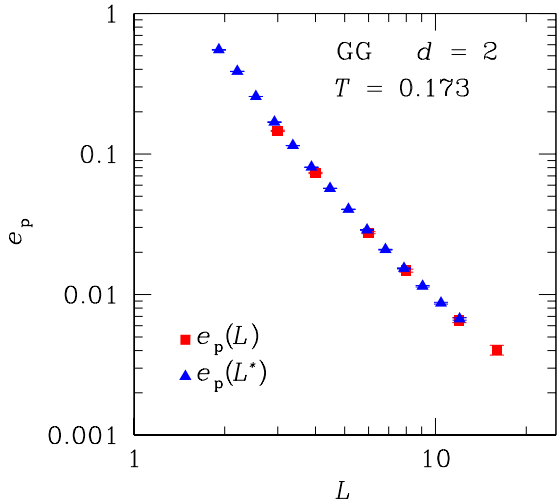}
\vspace*{-1.0cm}
\caption{(Color online)
Finite-size 
equilibrium energy differences $e_{\rm p}(L)$ [Eq.~(\ref{eq:ep2})] vs 
$L$ (squares) for the two-dimensional GG at $T = 0.173$.
On the same plot the energy differences $e_{\rm p}(t_{\rm w})$ at 
time $t_{\rm w}$ [Eq.~(\ref{eq:ep1})] after 
a quench are plotted against 
$L^{*}(t_{\rm w}) = A(t_{\rm w})t_{\rm w}^{1/z(T)}$ with identical 
parameters $A(T)$ and $z(T)$ as used in Fig.~\ref{fig:D2} (triangles). 
In this figure $e_{\rm p} = e + 1.502$.}
\label{fig:D3}
\end{figure}

\begin{figure}
\includegraphics[width=8.0cm]{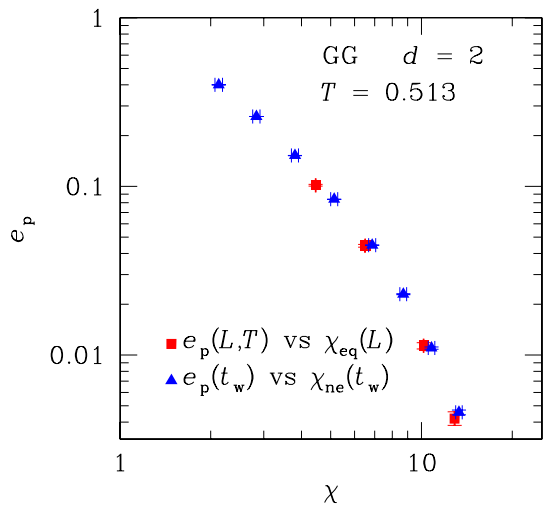}
\vspace*{-1.0cm}
\caption{(Color online)
Two-dimensional GG equilibrium (squares) and nonequilibrium (triangles) 
susceptibility and energy data for $T = 0.513$, as in 
Fig.~\ref{fig:D1}. Note the nontrivial functional form.
}
\label{fig:D4}
\end{figure}

\begin{figure}
\includegraphics[width=8.0cm]{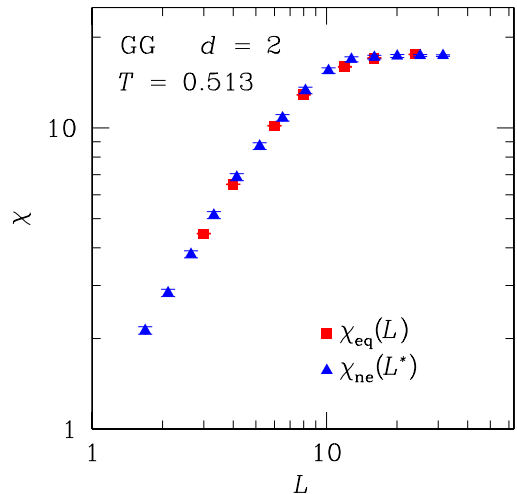}
\vspace*{-1.0cm}
\caption{(Color online)
Two-dimensional GG equilibrium (squares) and nonequilibrium 
(triangles) susceptibility data plotted as in Fig.~\ref{fig:D2}, 
for $T=0.513$ with appropriate adjustment for optimum $A(T)$ and $z(T)$ 
for this temperature.
In this figure $L^*(t_{\rm w}) = 1.346 t_{\rm w}^{1/3.08}$.}
\label{fig:D5}
\end{figure}

\begin{figure}
\includegraphics[width=8.0cm]{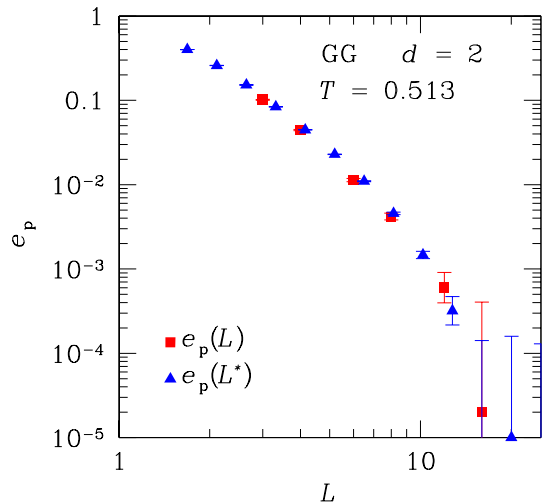}
\vspace*{-1.0cm}
\caption{(Color online)
Two-dimensional GG equilibrium (squares) and nonequilibrium 
(triangles) energy data are plotted as in Fig.~\ref{fig:D3}, 
for $T=0.513$ with the same parameters $A(T)$ and $z(T)$ used 
in Fig.~\ref{fig:D5}. In this figure $e_{\rm p} = e + 1.263$.}
\label{fig:D6}
\end{figure}

As the systems are always paramagnetic at finite $T$, for the 
whole temperature range $\chi_{\rm eq}(L)$ must finally 
saturate at the thermodynamic infinite-size limit and a plot of 
$\log[\chi_{\rm eq}(L,T)]$ against $\log(L)$ is always curved (although the 
curvature is weak at small sizes and low $T$). At each temperature a 
comparison between the nonequilibrium data for the susceptibility 
$\chi_{\rm ne}(t_{\rm w},T)$ and energy $e_{\rm ne}(t_{\rm w},T)$ 
and the equilibrium 
susceptibility $\chi_{\rm eq}(L,T)$ and energy $e_{\rm eq}(L,T)$ can be made. 
Figures \ref{fig:D1}--\ref{fig:D6} show examples at two different 
temperatures $T=0.173$ and $T=0.513$. In each case a plot is first made of 
\begin{equation}
e_{\rm p}(t_{\rm w})=[e_{\rm ne}(t_{\rm w},T) - 
e_{\rm eq}(\infty,T)]
\label{eq:ep1}
\end{equation}
against $\chi_{\rm ne}(t_{\rm w},T)$ and of 
\begin{equation}
e_{\rm p}(L,T)=[e_{\rm eq}(L,T) - e_{\rm eq}(\infty,T)] 
\label{eq:ep2}
\end{equation}
against 
$\chi_{\rm eq}(L,T)$. The 
excellent superpositions show that to good accuracy the effective length 
scale $L^{*}(t_{\rm w}, T)$ for the two observables is the same throughout the 
anneal at each temperature.
In both equations $e_{\rm eq}(\infty,T)$ is estimated by extrapolation. The
evaluation of $L^{*}(t_{\rm w}, T)$ from the energy data is not sensitive to
$e_{\rm eq}(\infty,T)$ as long as exactly the same value is used for 
the equilibrium and nonequilibrium data.

To obtain $L^{*}(t_{\rm w}, T)$ explicitly, the $\chi_{\rm ne}(t_{\rm w},T)$ 
data are 
scaled onto the equilibrium $\chi_{\rm eq}(L,T)$ data, translating to 
$\chi(L^{*},T)$ by assuming the space-time relation in Eq.~(\ref{eff_exponent})
and adjusting $A(T)$ and $z(T)$ to obtain optimum scaling. Scaling plots are 
shown for the same two temperatures. The condition $L^{*}(t_{\rm w},T) \ll 
L_{\rm b}$ 
is well satisfied for all the data with the present ranges of sizes and 
maximum anneal time. At the higher temperatures studied (as in Fig.~\ref{fig:D4}
for $T=0.513$), $\chi_{\rm eq}(L,T)$ and $\chi_{\rm ne}(t_{\rm w},T)$ 
saturate and the 
range of points from which the fit is usable is restricted to lengths and 
times before the onset of saturation. For $T=0.513$ accurate scaling can 
still be obtained, but this condition leads to a practical upper limit on 
the temperatures over which $A(T)$ and $z(T)$ can be estimated.
Note that these two temperatures are chosen as typical examples of 
the behavior in the two regimes in which the data sets do not or do 
arrive at saturation, respectively, within the available ranges of $L$ 
and $t_{\rm w}$. In other space dimensions and for the Ising systems
the equivalent plots always take on one or other of the behaviors 
depending on the temperature.


The size-dependent equilibrium energy data $e_{\rm p}(L,T)=[e_{\rm eq}(L,T) - 
e(\infty,T)]$ and the time-dependent nonequilibrium energy data 
$e_{\rm p}(t,T)=[e_{\rm ne}(t_{\rm w},T) - e(\infty,T)]$ 
translated to $e_{\rm p}(L^{*},T)$ 
can also be plotted together assuming exactly the same scaling as for the 
susceptibility data as discussed above. To the present accuracy, for these 
temperatures the $z(T)$ values obtained from the energy data are 
indistinguishable from the values estimated from the susceptibility scaling 
[although the effective prefactors $A(T)$ are slightly different].
The errors in each $z(T)$ data point are subjective estimates obtained 
by varying the parameters around the optimal values.

\begin{figure}
\includegraphics[width=8.0cm]{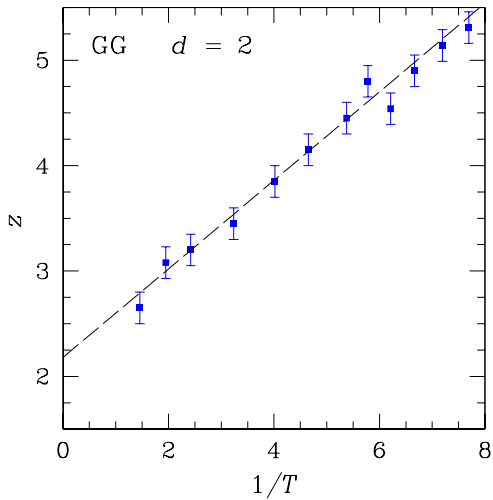}
\vspace*{-1.0cm}
\caption{(Color online)
Effective dynamical exponent $z(T)$ as a function
of $1/T$ for the two-dimensional GG. The dashed line corresponds to a fit
of the form $z(T) = 2.18(9) + 0.41(2)/T$ and is meant as a guide to the eye.
}
\label{fig:D7}
\end{figure}

The estimates for $z(T)$ against $1/T$ are shown in Fig.~\ref{fig:D7}. 
The data can be parametrized using
\begin{equation}
z(T) =  2.18(9)+0.41(2)/T \; ,
\end{equation}
with $\chi^2 \simeq 0.60$ for the fit. This implies a 
diverging dynamical exponent $z(T)$ as $T$ approaches zero
and $z(T)$ tending to near $2$ at 
infinite temperature. Note that $z = 2$ for a random walk.

We conclude from this section that for the GG in dimension 2 which orders 
at zero temperature, a time-dependent length scale $L^{*}(t_{\rm w},T)$ can be 
measured over a wide range of temperatures in the paramagnetic state. This 
length scale obeys the effective dynamical critical exponent scaling 
relationship, 
Eq.~(\ref{eff_exponent}). The sets of values of $z(T)$ determined from the 
nonequilibrium spin-glass susceptibility $\chi_{\rm ne}(t_{\rm w},T)$ and the 
nonequilibrium energy per spin $e_{\rm ne}(t_{\rm w},T)$ are the same to within 
the present precision.

\subsection{Three dimensions}
\label{sec:3dgg}

The GG in dimension 3 has an ordering temperature 
$T_{\rm c}=0.46(1)$.\cite{olson:00,katzgraber:04f} 
The analysis protocol used is essentially the 
same as in dimension 2 (Sec.~\ref{sec:2dgg}); parameters of the simulation are
listed in the Appendix, Table \ref{tab_gg_eq_3} 
for equilibrium and Table \ref{tab_gg_neq} for
nonequilibrium measurements, respectively.
Below $T_{\rm c}$ and in the $L$ and $t_{\rm w}$ ranges that 
we have studied, the equilibrium susceptibility increases as 
$\chi_{\rm eq}(L,T) \propto L^{x(T)}$ and the nonequilibrium susceptibility 
as $\chi_{\rm ne}(t_{\rm w},T) \propto t_{\rm w}^{y(T)}$ with 
$T$-dependent $x(T)$ and $y(T)$. This algebraic behavior 
facilitates the analysis because the log-log plots of $\chi$ are all 
straight lines.

As in the two-dimensional case, at each temperature an 
effective dynamical exponent 
$z(T)$ and prefactor $A(T)$ can be defined from the scaling of the 
equilibrium and nonequilibrium susceptibility data using
Eq.~(\ref{eff_exponent}). As in two dimensions the energy 
data can be scaled 
satisfactorily using the same $z(T)$ obtained at each $T$ from the 
analysis of the susceptibility data. 
Below $T_{\rm c}$ the prefactors $A(T)$ are slightly different 
for both the energy, as well as the scaling of the susceptibility (not shown).

In three dimensions $z(T)$ also increases $\sim 1/T$ and has an intercept 
$z(\infty)\approx 2$---i.e., 
\begin{equation}
z(T)= 1.95(8)+1.17(4)/T\; , 
\end{equation}
with $\chi^2 \simeq 1.06$ for the fit.

The present estimate for 
$z_{\rm c}$ at $T_{\rm c} =0.46(1)$ is $z_{\rm c}= 4.5(1)$, 
in agreement with previous estimates.\cite{katzgraber:04f}

\begin{figure}
\includegraphics[width=8.0cm]{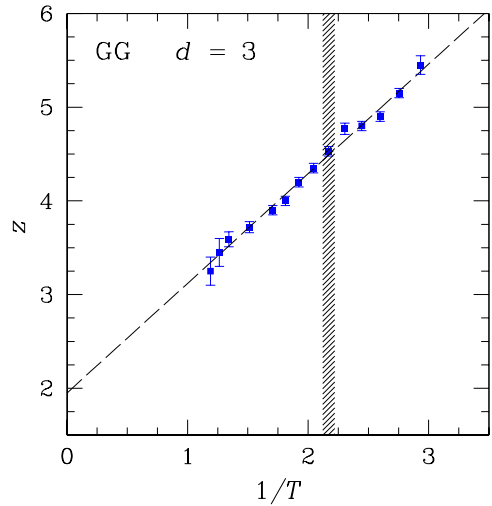}
\vspace*{-1.0cm}
\caption{(Color online)
Effective dynamical exponent $z(T)$ as a 
function of $1/T$ for the three-dimensional GG. The dashed 
line corresponds to a 
fit of the form $z(T) = 1.95(8) + 1.17(4)/T$. The shaded region denotes 
$T_{\rm c} = 0.46(1)$.
In this and the following figures, the width of
the shaded area represents the error bar on the estimate of the critical
temperature.}
\label{fig:D8}
\end{figure}

It is important to note that $z(T)$ traverses $T_{\rm c}$ smoothly with no 
apparent anomaly; see Fig.~\ref{fig:D8}.  
This implies that in terms of the evolution 
of length scales with time, the dynamics above and below the ordering 
temperature follow the same pattern although in the final equilibrium 
configuration all the spins are ordered below $T_{\rm c}$ and while they are
only correlated over a finite
length scale above $T_{\rm c}$. 
In this system the critical behavior is not exceptional as far as the length 
scale dynamics is concerned.

\subsection{Four dimensions}
\label{sec:gg4d}

The GG in four dimensions has an ordering transition at 
$T \approx 0.89(1)$.\cite{katzgraber:04f} 
We perform a similar analysis as done in the two- and three-dimensional case.
Parameters of the simulation are listed in the Appendix, Tables
\ref{tab_gg_eq_4} and \ref{tab_gg_neq}.
A very similar pattern of behavior is 
observed as in the GG with lower space dimensions, with $z(T)$ increasing 
linearly with inverse temperature---i.e.,
\begin{equation}
z(T)= 1.75(13)+2.4(1)/T \; . 
\end{equation}
Here $\chi^2 \simeq 0.3$ for the fit.

\begin{figure}
\includegraphics[width=8.0cm]{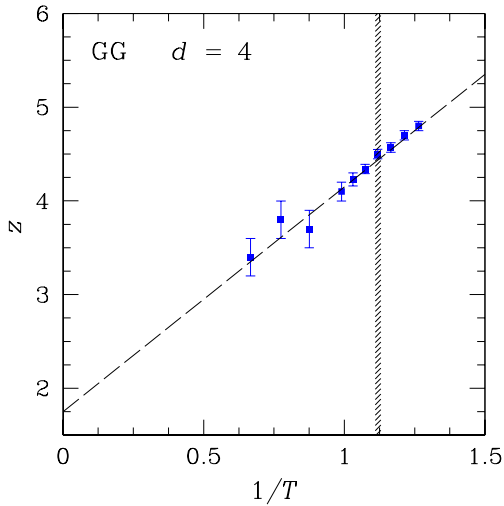}
\vspace*{-1.0cm}
\caption{(Color online)
Effective dynamical exponent $z(T)$ as a
function of $1/T$ for
the four-dimensional GG.
The dashed line corresponds to a fit of the form 
$z(T) = 1.75(13) + 2.4(1)/T$.
The shaded region denotes
$T_{\rm c} = 0.89(1)$.}
\label{fig:D9}
\end{figure}

As in dimension 3 there is no sign of any change of behavior in the region of 
$T_{\rm c}$; see Fig.~\ref{fig:D9}. 
At $T_{\rm c} \approx 0.89$ we obtain $z_{\rm c}=4.5(1)$, in 
agreement with Ref.~\onlinecite{katzgraber:04f}.

\section{Ising Spin Glass}
\label{sec:isg}

The Hamiltonian of the Edwards-Anderson Ising spin glass\cite{edwards:75} is
given by
\begin{equation}
{\cal H} = - \sum_{\langle i,j\rangle} J_{ij} S_i S_j ,
\label{eq:ham}
\end{equation}
where the sum is over nearest-neighbor pairs of sites on a hypercubic lattice
in $d$ dimensions,
the $S_i$ are Ising spins taking values $\pm 1$, and the $J_{ij}$ are Gaussian
distributed with zero mean and standard deviation unity. Simulations are done
using periodic boundary conditions. Parameters of the simulation are listed in
the Appendix.

\subsection{Two dimensions}
\label{sec:2disg}

As in the case of the two-dimensional GG, 
the EA ISG with Gaussian interactions in 
dimension 2 orders only at zero 
temperature;\cite{mcmillan:84b,bray:84,rieger:96,palassini:99a,hartmann:01a,hartmann:02c} 
thus, all the data concern the 
paramagnetic regime. Details of the simulations are summarized in the
Appendix, Tables \ref{tab_isg_eq_2} and \ref{tab_isg_neq}.
The data show that an analysis according to Eq.~(\ref{eff_exponent})
provides an excellent parametrization 
of the growth of correlations. The results are in good agreement with 
direct measurements of the correlation functions.\cite{kisker:96,rieger:04}

The energy per spin data, $e(t_{\rm w},T)$, can be parametrized consistently in 
terms of the same effective set of dynamical exponents $z(T)$ 
as used for the analysis of the $\chi(T)$ data.

\begin{figure}
\includegraphics[width=8.0cm]{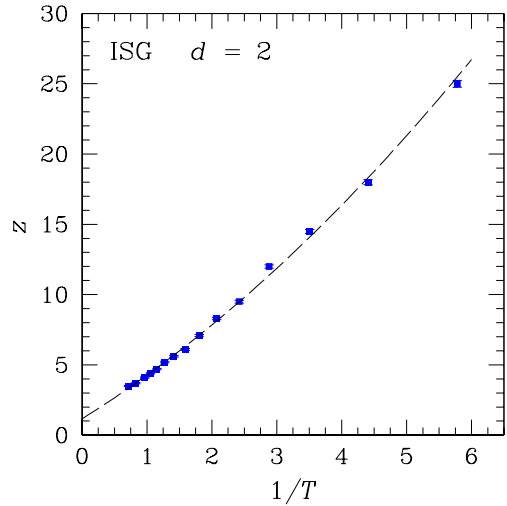}
\vspace*{-1.0cm}
\caption{(Color online)
The effective dynamical exponent $z(T)$ as a function of $1/T$ for
the two-dimensional ISG. The line is a guide to the eye.}
\label{fig:D10}
\end{figure}

The temperature dependence of $z(T)$ is much stronger than for the GG 
systems, and $z(T)$ deviates somewhat from a linear variation with inverse 
temperature; cf.~Fig.~\ref{fig:D10}. 
For $1/T \leq 5$, the $z(T)$ data can be 
approximately parametrized by $z(T) \simeq 3.9/T$. (It is possible that 
the curve could bend at low $1/T$ to a nonzero intercept.)
The present analysis is consistent with those of 
Refs.~\onlinecite{kisker:96} and \onlinecite{rieger:04}, 
who obtain $z(T) \simeq 4.3/T$ using a different numerical 
technique. In Fig.~\ref{fig:D10} the dashed line is a quadratic fit 
in $x = 1/T$ and should serve as a guide to the eye.

\subsection{Three dimensions}
\label{sec:3disg}

There is general consensus that the freezing temperature of the 
three-dimensional EA 
ISG with Gaussian interactions is 
$T_{\rm c} = 0.94(2)$ (Refs.~\onlinecite{marinari:98,campbell:00,ballesteros:00,mari:01})
and that the dynamical critical
exponent is $z_{\rm c} \simeq 6.45(10)$. The equilibrium SG susceptibility 
$\chi_{\rm eq}(L,T)$ and energy per spin, $e_{\rm eq}(L,T)$, are measured at 
temperatures between $T=0.1$ and $T=3.0$ for several intermediate sample
sizes, see Table \ref{tab_isg_eq_3}.
Nonequilibrium measurements are made for $L_{\rm b} = 24$; see Table
\ref{tab_isg_neq}.
For temperatures up to $T_{\rm c}$ and for the range of sizes used, to a 
good approximation (as in the GG below $T_{\rm c}$)
\begin{equation}
\chi_{\rm eq}(L,T) = C(T)L^{x(T)} \; ,
\label{chiL_eqn}
\end{equation}
with $C(T)$ a temperature-dependent prefactor and $x(T)$ an effective 
exponent which becomes equal to the true static critical exponent $2-\eta$ 
at $T_{\rm c}$.  $x(T)$ can never exceed $d$; for this system it reaches values 
very close to $3$ as $T$ tends to zero. $C(T)$ remains very close to $1$ 
for the whole temperature range. We can note that by definition 
$\chi_{\rm eq}(L=1)=1$ for all $T$. In the low-temperature range, for the 
equilibrium susceptibility $\chi_{\rm eq}(L,T)$ we make the assumption that 
$\log[\chi_{\rm eq}(L,T)]$ extrapolates linearly with $\log(L)$ to zero at 
$L=1$. ($L=2$ measurements have not been used, as at this particular size there 
are intrinsic ``wrap-around'' problems associated with the definition of the 
interactions.)

\begin{figure}
\includegraphics[width=8.0cm]{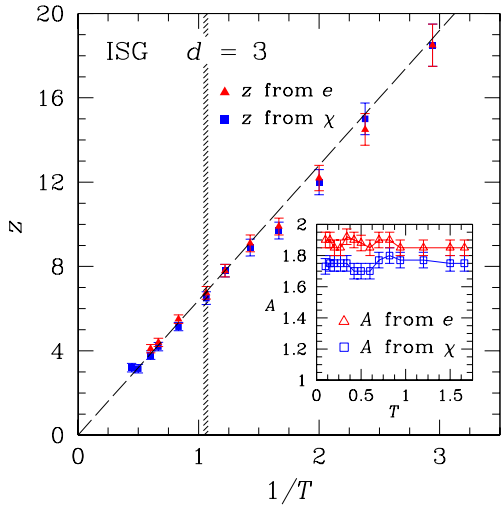}
\vspace*{-1.0cm}
\caption{(Color online)
The effective dynamical exponent $z(T)$ as a function of $1/T$ for
the three-dimensional ISG. The line corresponds to a fit of the form
$z(T) = 6.40(15)/T$. The squares correspond to data for $z(T)$ as estimated from
the energy, whereas the triangles represent data for $z(T)$ calculated using
the spin-glass susceptibility. The inset shows the prefactor $A$ of the
scaling relation in Eq.~(\ref{eff_exponent}) as estimated for the energy and
susceptibility. The data show little temperature dependence.
}
\label{fig:D11}
\end{figure}

The effective dynamical critical exponent scaling parametrization analysis 
using Eq.~(\ref{eff_exponent}) is satisfactory over the whole temperature 
range covered. Below $T_{\rm c}$, the present $z(T)$ values, 
Fig.~\ref{fig:D11}, are 
consistent with those of Refs.~\onlinecite{parisi:96} and 
\onlinecite{kisker:96} but more accurate. 
The prefactors $A(T)$ vary only slightly with $T$, and $A(T) 
\simeq 2B(T)$ where $B(T)$ are the prefactors for the coherence length 
$\ell(t_{\rm w},T)$ estimated by Refs.~\onlinecite{parisi:96} and
\onlinecite{kisker:96}. 
This agreement 
confirms the conjecture made above that as a general rule the correlation 
length $\ell(t_{\rm w})$ can be taken as equal to $L^{*}(t_{\rm w})/2$ to a 
good approximation. 
For temperatures above $T_{\rm c}$, a scaling of the effective 
exponent form remains very satisfactory; see Fig.~\ref{fig:D11}.

As found by Refs.~\onlinecite{parisi:96} and \onlinecite{kisker:96}, 
$z(T)$ varies approximately 
linearly with $1/T$. The present data (which are more accurate than those 
of the previous work), including points in the paramagnetic region up to 
about $2.5 T_{\rm c}$, are consistent with $z(T) \simeq 6.40(15)/T$
($\chi^2 \simeq 0.85$).
The points for $z(T)$ vary smoothly and continuously through $T_{\rm c}$, as in 
the three- and four-dimensional GG.

Again, one can carry out a scaling plot for the nonequilibrium energy 
$e_{\rm ne}(t_{\rm w},T)$ in the same way as for the susceptibility. The effective 
$z(T)$ values estimated from the energy scaling are consistent with the 
values from the susceptibility scaling, but the prefactors $A(T)$ become 
slightly different in the lower temperature range.
This confirms that one single anneal-time-dependent length growth law 
controls both susceptibility and energy during the anneal, which seems a 
more satisfactory form of analysis than, for instance, that given in 
Sec.~VI A of Ref.~\onlinecite{yoshino:02}.

It would be of interest to carry out further measurements so as to obtain 
significantly more information at low and moderate temperatures, but this 
would require nonequilibrium simulations extending to much longer 
annealing times.

\subsection{Four dimensions}
\label{sec:4disg}

For the EA ISG with Gaussian couplings in four dimensions it is known that 
$T_{\rm c}=1.78(2)$,\cite{parisi:96,bernardi:97,campbell:00} 
with a dynamical critical exponent 
$z_{\rm c}=4.9(2)$.\cite{bernardi:97,campbell:00} The present data are
analyzed in just the same way as for the three-dimensional ISG. 
There is excellent 
agreement between estimates for $z(T)$ between susceptibility and energy 
data.

\begin{figure}
\includegraphics[width=8.0cm]{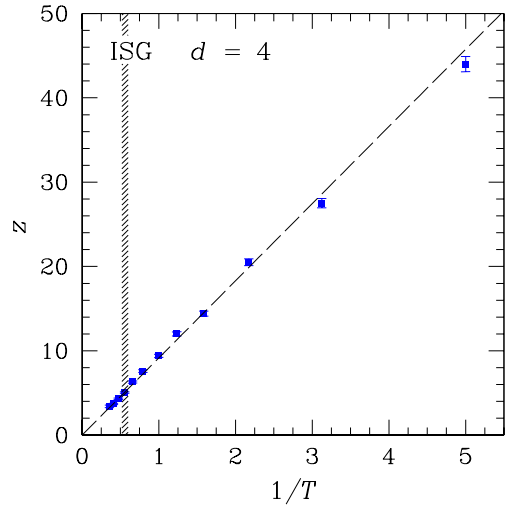}
\vspace*{-1.0cm}
\caption{(Color online)
The effective dynamical exponent $z(T)$ as a function of $1/T$ for
the four-dimensional ISG. The line corresponds to a fit of the form 
$z(T) = 9.15(20)/T$.
}
\label{fig:D12}
\end{figure}
                                                                                
In four dimensions $z(T)$ varies approximately linearly with inverse 
temperature---i.e., $z(t) \simeq 9.15(20)/T$ with $\chi^2 \simeq 2.8$; 
see Fig.~\ref{fig:D12}. 
The estimate for $z$ at $T_{\rm c}$ is $z_{\rm c} = 5.1(2)$, in 
agreement with previous work.\cite{bernardi:97} Once more, $z(T)$ evolves 
smoothly as $T$ passes through $T_{\rm c}$.

\section{Summary}
\label{sec:summary}

The observed behavior of the dynamical exponent $z(T)$
for the six systems studied is summarized in Table \ref{tab:z_res}.
It can be seen that for the GG systems the data in 
each dimension can be parametrized in the form $z(T)\simeq 
a+b/T$ with the constants $a \approx 2$ and $b$ increasing with 
dimension $d$. For the EA ISG systems, $z(T)\simeq b/T$---i.e., $a = 0$, with 
$b$ values again increasing strongly with space dimension $d$; 
in each dimension 
the value of $b$ is higher than that of the GG by a factor of 5--10. 
For each of the ISG's the data could be compatible with a nonzero 
infinite-temperature intercept if there is some curvature in $z(1/T)$ at 
temperatures higher than those we have studied. In two dimensions the data
potentially suggest such a behavior.

\begin{table}
\caption{
\label{tab:z_res}
Temperature dependence of the dynamical exponent 
$z(T) = a + b/T$ for the
different models studied at different space dimensions $d$. GG refers to the
gauge glass, ISG to the Ising spin glass with Gaussian interactions.
For the Ising spin glass in two dimensions the quoted value of $b$
corresponds to a linear fit which is a poor approximation to the behavior seen
in this system (marked with an asterisk). 
}
\begin{tabular*}{\columnwidth}{@{\extracolsep{\fill}} l l l l  }
\hline
\hline
System & $d$ & $a$        & $b$  \\
\hline
GG         & $2$ & $2.18(9)$  & $0.41(2)$  \\
GG         & $3$ & $1.95(8)$  & $1.17(4)$  \\
GG         & $4$ & $1.75(13)$ & $2.4(1)$   \\
ISG$^\ast$ & $2$ & $0$        & $3.9(1)$   \\
ISG        & $3$ & $0$        & $6.40(15)$ \\
ISG        & $4$ & $0$        & $9.15(20)$ \\
\hline
\hline
\end{tabular*}
\end{table}

\section{Droplet dynamics approach}
\label{sec:droplets}

We have obtained very satisfactory scaling of the Ising spin glass and 
gauge glass dynamics using the effective exponent 
parametrization, Eq.~(\ref{eff_exponent}). We now discuss an 
alternative analysis based on the dynamical droplet approach, which 
distinguishes critical behavior near $T_{\rm c}$ from activated behavior with 
a barrier exponent $\psi$ further below $T_{\rm c}$.\cite{bouchaud:02a} 
Berthier and Bouchaud\cite{berthier:02,berthier:03a} (BB) carried out 
four-spin correlation function measurements on the three-dimensional 
EA ISG similar to those 
of Refs.~\onlinecite{parisi:96} and \onlinecite{kisker:96}, 
but analyzed their data on the basis of 
the droplet dynamics formula.\cite{bouchaud:02a}

A similar analysis of Heisenberg spin-glass dynamics (see Fig.~11 of 
Ref.~\onlinecite{berthier:04}) shows qualitatively different behavior 
from that 
of the systems reported here. We have no data on Heisenberg systems 
although the present technique could potentially be applied there as well.

The BB droplet dynamics leads to a growth law with time corresponding to a 
coherence length $\ell(t)$
\begin{equation}
t_{\rm w}[\ell(t,T)] \simeq 
\tau^{*}[\ell(t,T)]^{z_{\rm c}}\exp[\Upsilon(T)\ell(t,T)^{\psi}/T] \; ,
\label{BB_eqn}
\end{equation}
where $z_{\rm c}$ is the dynamical critical exponent, $\psi$ is the barrier 
exponent, $\Upsilon(T) =\Upsilon_0(1-T/T_{\rm c})^{\nu\psi}$, and $\tau^{*}$ is 
a prefactor. BB scale the data using $z_{\rm c} = 7$, $\nu = 1.65$, 
$\psi = 1$, and $\Upsilon_0 = 5.5$ and assume $\tau^{*} \sim 2$ but to find 
good overall fits they needed to choose temperature-dependent values for 
$\tau^{*}$.

Yoshino {\em et al.}\cite{yoshino:02} made a closely related analysis of 
measurements on the four-dimensional $\pm J$ ISG. 
They introduce a ``crossover length''
\begin{equation}
L_0(T)= T^{1/\psi}(1-T/T_{\rm c})^{-\nu}
\label{eq:lnot}
\end{equation} 
and a ``crossover time''
\begin{equation}
\tau_0(T) = [L_0(T)]^{z_{\rm c}}
\label{eq:tnot}
\end{equation}
together with a general scaling law for the coherence length,
\begin{equation}
\ell(t_{\rm w})/L_0(T) = f_{\ell}[t_{\rm w}/\tau_0(T)]\; ,
\label{yoshino_eqn}
\end{equation}
which relates the coherence length $\ell(t_{\rm w})$ to the elapsed time 
$t_{\rm w}$. Here $f_{\ell}$ represents an unknown scaling function. 
The scaling curve depends on the critical parameters ($T_{\rm c}$, $z_{\rm c}$, 
and $\nu$) 
which can be taken as known from measurements at $T_{\rm c}$, 
and on the barrier 
exponent $\psi$. It can be readily shown that the BB equation, 
Eq.~(\ref{BB_eqn}), is an inversion of the scaling equation of Yoshino {\it 
et al}., Eq.~(\ref{yoshino_eqn}), with an explicit functional form which can 
be written as
\begin{equation}
t_{\rm w}/\tau_0(T) = 
	\tau^{*}[\ell (t_{\rm w})/L_0(T)]^{z_{\rm c}}
	\exp\{\Upsilon_0[\ell (t_{\rm w})/L_0(T)]^{\psi}\} \;.
\label{BB_scaling}
\end{equation}

\begin{figure}
\includegraphics[width=8.0cm]{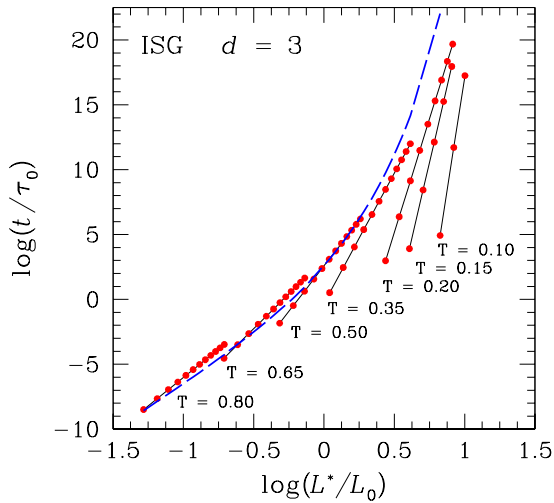}
\vspace*{-1.0cm}
\caption{(Color online)
Three-dimensional ISG data plotted following to the parametrization
of Yoshino {\it et al.} (Ref.~\onlinecite{yoshino:02}). 
According to this droplet
dynamics approach all the data presented in this way should lie on a
single scaling curve. The long-dashed line is the particular scaling curve
calculated using the specific parameters proposed by Berthier and Bouchaud
in Ref.~\onlinecite{berthier:02} for this system.
For the definitions of the parameters
$L_0$ and $\tau_0$ see the text, Eqs.~(\ref{eq:lnot}) and (\ref{eq:tnot}),
respectively.}
\label{fig:D13}
\end{figure}

We now analyze the present data for the EA ISG in three dimensions
in terms of the 
BB/Yoshino droplet dynamics approach.\cite{comment:4d}
In Fig.~\ref{fig:D13} we plot the present 
data in three dimensions for $L^{*}(t_{\rm w},T)$ 
as a function of $t_{\rm w}$ at different $T$ in the 
scaling form proposed by Yoshino {\it et al.};\cite{yoshino:02} i.e., we 
plot $L^{*}(t_{\rm w},T)/2L_0(T)$ against $t_{\rm w}/\tau_0(T)$ with $L_0(T)= 
T^{1/\psi}(1-T/T_{\rm c})^{-\nu}$ and $\tau_0 = [L_0(T)]^{z_{\rm c}}$ 
[the factor $1/2$ corresponds to the translation 
from $L^{*}(t_{\rm w},T)$ to $\ell(t_{\rm w},T)$] 
(see inset of Fig.~\ref{fig:D11}).
For this scaling we adopt for the three-dimensional 
EA system the parameters proposed by BB:
$z_{\rm c} = 7$, $\nu = 1.65$, and $\psi = 1$. 
According to Yoshino {\it et al.}, 
the entire data set scaled this way should lie on a single (but 
unspecified) scaling curve. The extra BB scaling parameters define one 
specific scaling curve within the same plot. On the ``Yoshino plot'' we 
therefore draw the full BB scaling plot, Eq.~(\ref{BB_scaling}), using the 
two remaining fit parameters from the BB parameter set: $\Upsilon_0 = 
5.5$ and $\tau^{*}= 2$.

In the temperature region from $T_{\rm c}$ down to about $T_{\rm c}/2$ 
the overall 
Yoshino scaling and the agreement between the present data and the BB 
scaling curve is acceptable. This is the same temperature range as covered 
by the BB simulations. However, in the temperature range below $T_{\rm c}/2$ 
where the activated droplet dynamics should be valid because little 
is affected by critical dynamics, the scaling breaks down. The 
curves for different $T$ are not superimposed, and the deviations from the 
BB curve correspond to many orders of magnitude along the time axis.

An analysis of the EA ISG data in four dimensions
leads to the same conclusion. The 
measurements of Yoshino {\it et al.}\cite{yoshino:02} were made on the 
$\pm J$ model while the present measurements correspond to the model
with Gaussian disorder, but if we adopt the critical parameters 
$T_{\rm c}=1.8$, 
$z_{\rm c}=4.5$, and $\nu=0.93$, together with $\psi=4.5$ from Yoshino 
{\it et al.},\cite{yoshino:02} we obtain an overall scaling plot for 
the present EA ISG results in four dimensions. Once 
again the agreement between the scaling prediction 
(a unique scaling curve) and the data set is acceptable for 
temperatures $T_{\rm c}/2 \leq T \leq T_{\rm c}$, 
but it breaks down at lower temperatures (not shown).

Thus for these two canonical ISG systems the standard droplet dynamics 
scaling below $T_{\rm c}$ does not give a satisfactory global account of the 
data; by suitably adjusting the various fit parameters the droplet scaling 
relation, Eq.~(\ref{yoshino_eqn}), can be made to conform reasonably well to 
the data over limited ranges of temperature and annealing time, but at low 
temperatures this form of scaling predicts a time scale $t/\tau_{0}$
which diverges much more rapidly with increasing length scale $L^{*}/L_{0}$
than is observed.

If we attempt to determine $\psi$ for the three-dimensional GG
from scaling our data 
using this droplet expression, we find only poor scaling 
whatever $\psi$ values are assumed for a very wide range of values.

Kisker {\em et al.},\cite{kisker:96}~as well as fitting very successfully their 
EA ISG data in two and three dimensions 
to Eq.~(\ref{eff_exponent}), also compared their data 
to a different phenomenological droplet-dynamics-inspired formula:
\begin{equation}
\ell(t_{\rm w})-\ell_0 = \lambda(T)[\ln(t_{\rm w})]^{1/\psi}
\label{kisker_eqn}
\end{equation}
and found that this relation was capable of giving a satisfactory fit to 
their low-temperature data. However, as they point out in 
Ref.~\onlinecite{kisker:96}, this formulation has a number of defects; in 
particular, one would not expect the droplet formula to be applicable in 
the paramagnetic region. We can also examine the mathematical basis of the 
equation. Suppose that the true time dependence is 
$\ell(t_{\rm w})=At_{\rm w}^{1/z}$---i.e., 
$\ln[\ell(t_{\rm w})]=\ln(A)+(1/z)\ln(t_{\rm w})$---with $z$ 
large. The parametrization, Eq.~(\ref{kisker_eqn}), can be recast as 
$\ell(t_{\rm w})=\ell_{0}\cdot(1+x)$ or $\ln[\ell(t_{\rm w})]= \ln(\ell_0)+ 
\ln(1+x)$, 
with 
$x=[B\ln(t_{\rm w})]^{1/\psi}$, where $B=(\lambda(T)/\ell_0)^{\psi}$ is small. 
Because $\ln(1+x)=x-x^2/2+...$, we have 
$\ln[\ell(t_{\rm w})]=\ln(\ell_0)+ [B\ln(t_{\rm w})]^{1/\psi}-
[B\ln(t_{\rm w})]^{2/\psi}/2+\cdots$.
If $1/\psi > 1$, on a $\log[\ell(t_{\rm w})]$ against $\log(t_{\rm w})$ 
plot there will 
be an upbending from the second term compensated by a growing negative 
third term. Trial and error shows that if $\psi \simeq 2/3$, this 
parametrization produces to quite good accuracy a linear dependence of 
$\log[\ell(t_{\rm w})]$ on $\log(t_{\rm w})$ 
over a wide range of $t_{\rm w}$, typically from 
$t_{\rm w}=10$ to $t_{\rm w}=10^8$, successfully mimicking 
$\ln[\ell(t_{\rm w})]=\ln(A)+(1/z)\ln(t_{\rm w})$. Thus a fit to the 
parametrization in Eq.~(\ref{kisker_eqn}) can be obtained with a 
pseudo ``barrier exponent'' $\psi $ even when the true behavior is 
better described by 
$\ell(t_{\rm w})=At_{\rm w}^{1/z}$, so this fit is a mathematical artifact. The 
value of 
$\psi$ obtained from such a fit can be expected to be $\simeq 2/3$ 
regardless of  the system studied, which explains why the apparent 
$\psi$ estimated in 
Ref.~\onlinecite{kisker:96} both for the three-dimensional ISG 
well below $T_{\rm c}$ and for the 
paramagnetic two-dimensional ISG at low temperatures are close to this value.

We conclude from this section that the standard droplet scaling approach
does not give a satisfactory overall account of the numerical Ising 
spin-glass dynamical data in dimensions 3 and 4.

\section{Conclusion}
\label{sec:concl}

We have studied the dynamical behavior of ISG's and GG's in dimensions 2, 3, 
and 4 as functions of anneal time $t_{\rm w}$ after a quench to a 
temperature $T$. 
We define a time-dependent correlation length scale 
$L^{*}(t_{\rm w},T)$ by relating the 
time-dependent nonequilibrium susceptibility 
$\chi_{\rm ne}(t_{\rm w},T)$ to the size-dependent equilibrium susceptibility 
$\chi_{\rm eq}(L,T)$ and the time-dependent energy $e_{\rm ne}(t_{\rm w},T)$ 
data to 
the equilibrium energy $e_{\rm eq}(L,T)$ data. In each system and at all 
temperatures (below and above the freezing temperature $T_{\rm c}$,
as well as at 
$T_{\rm c}$) a growth law for the length scale $L^{*}(t_{\rm w},T)=
A(T)t_{\rm w}^{1/z(T)}$ 
gives a good parametrization of the data, with an effective 
temperature-dependent dynamical exponent $z(T)$ and a 
prefactor $A(T)$ which is only 
weakly temperature dependent. In each case, independent estimates of $z(T)$ 
from the susceptibility and from the energy measurements are the same 
within the precision of the measurements; the prefactors $A(T)$ 
corresponding to the susceptibility and to the energy can be marginally 
different. $z(T)$ diverges approximately as $1/T$ at low temperatures in 
all the systems, and at high temperatures it appears to tend to a 
limiting value consistent with $z(\infty) \sim 2$ for the GG systems and 
to a value which extrapolation suggests is close to zero for the ISG 
systems. The evolution of $z(T)$ as the temperature passes through the 
freezing temperature $T_{\rm c}$ is smooth for all four systems having a 
nonzero $T_{\rm c}$. In the Ising spin glasses the dynamic droplet 
critical scaling approach\cite{berthier:02,yoshino:02} is incompatible 
with the present measurements for the region below $T_{\rm c}/2$. The droplet 
picture gives no predictions whatsoever concerning time-length relationships
in the paramagnetic region above $T_{\rm c}$. The data show that the standard 
droplet dynamics 
assumptions\cite{fisher:86,fisher:88,fisher:88b,bouchaud:02a} 
of an effective barrier 
height increasing as a power of $\ell$ and disappearing at $T_{\rm c}$ are 
inappropriate for the systems studied.

Many recent experimental measurements of dynamics in spin 
glasses\cite{dupuis:01,bouchaud:02a,jonsson:02a,bert:04} have been interpreted 
using the droplet dynamics parametrization. It is nontrivial to transpose 
conclusions obtained from numerical data to the experimental regime; the 
effective time scales for experiments are vastly greater than for the 
numerical results, 
and length scales are slightly larger. It must also be kept in 
mind that there appears to be no way in which to measure spin-glass 
coherence lengths directly from experiment. As the present results 
invalidate the standard droplet dynamics in Ising spin glasses at least on 
time scales attainable numerically, the question should be raised as to 
what physical significance, if any, can be ascribed to parameters such as 
the barrier exponent $\psi$ which are obtained from scaling analyses of 
experimental data based on standard droplet dynamics.  As has been pointed 
out in Ref.~\onlinecite{jonsson:04}, for one canonical ISG system the range of 
published estimates of $\psi$ from different experiments extends from 
$0.03$ to 
$1.9$,\cite{mattsson:95,dupuis:01,bouchaud:02a,jonsson:02a,bert:04} which also 
suggests that the standard droplet dynamics parametrization is 
inappropriate. Recently, a rigorous bound for the barrier exponent inferred
from two-dimensional calculations\cite{amoruso:05} has been given 
with $0.25 < \psi < 0.54$.
Still, it would seem well worth attempting to review the 
experimental data to see if they can be reinterpreted in terms of an 
effective exponent length scale growth scenario compatible with the 
present numerical results.

The effective dynamical critical exponent scaling scenario 
with $z(T) \propto 1/T$ which provides a
satisfactory parametrization for the ISG systems is very similar to that
observed by Paul {\em et al.}\cite{paul:04,paul:05}~for random-bond
Ising ferromagnets (RBIM's) and diluted ferromagnets in dimension 2.
The physical mechanism for the time dependence of the correlation length in
these ferromagnetic models is domain coarsening, and the data are
interpreted in terms of effective barriers to domain growth which increase
logarithmically with size.\cite{rieger:93}
It would seem very plausible to ascribe
the correlation length growth in spin glasses to an analogous
mechanism. Equation (9) of Paul {\em et al.}\cite{paul:04,paul:05}~can 
be written as $z(T) = 2 + c/T$, $c$ a constant, in our terminology, 
which is precisely what we have observed
empirically for the GG systems. However, it should be noted that at the
ordering temperature $T_{\rm c}$ for the RBIM the effective dynamical 
exponent tends to the pure system value $z \sim 2$.\cite{paul:04} 
In the ISG's and GG's the dynamical critical exponent $z_{\rm c}$ 
is always much higher than $2$ and $z(T)$ continues
to decrease regularly as the temperature is raised through the
paramagnetic regime. This is true both above the ordering temperature when
$T_{\rm c}$ is nonzero or at all temperatures when  $T_{\rm c}=0$ 
which is the case in dimension 2.

The standard approach to the dynamics of the growth of the coherence 
length $\ell(t_{\rm w})$ with anneal time after quench $t_{\rm w}$ is to
assume that there are three principal relaxation regimes: paramagnetic at
temperatures well above $T_{\rm c}$, critical in the region around
$T_{\rm c}$, and activated at temperatures well below $T_{\rm c}$, 
each regime having a qualitatively different relaxation behavior. 
The present data show that for Ising spin glasses and for the gauge glass, 
a dynamical relationship of the standard critical form, 
$\ell(t_{\rm w}) = A(T)t_{\rm w}^{1/z(T)}$
with a temperature-dependent dynamical exponent $z(T)$, 
gives a good account of the nonequilibrium dynamics at each temperature 
and not only at $T_{\rm c}$. The data indicate that $z(T)$ varies 
smoothly as a function of temperature when passing through $T_{\rm c}$.

\begin{acknowledgments}

We would like to thank Ludovic Berthier for helpful
comments. The two-dimensional Ising spin-glass data have been taken from a
previous study (Ref.~\onlinecite{katzgraber:04}) 
done in collaboration with L.~W.~Lee and
A.~P.~Young. Part of the simulations were performed on the Asgard and Hreidar
clusters at ETH Z\"urich.
\end{acknowledgments}

\appendix*

\section{Numerical Details}

Equilibrium measurements are carried out with samples fully thermalized
using the exchange Monte Carlo (parallel tempering)
technique.\cite{hukushima:96,marinari:96} 
To ensure that the system is equilibrated, we perform a logarithmic data
binning of all observables (energy and spin-glass susceptibility) 
and require that the last three bins logarithmically spaced agree within
error bars and be independent of the number of Monte Carlo sweeps (MCS)
$N_{\mathrm{sweep}}$. In the case of the Ising spin glass we use the 
equilibration test for short-range spin glasses first introduced in
Ref.~\onlinecite{katzgraber:01}. Details about the equilibrium simulations for
the gauge glass are summarized in Tables \ref{tab_gg_eq_2},
\ref{tab_gg_eq_3}, and \ref{tab_gg_eq_4}, for $d = 2$, $3$, and $4$,
respectively. Details about the equilibrium simulations for the Ising spin
glass are presented in Tables \ref{tab_isg_eq_2},
\ref{tab_isg_eq_3}, and \ref{tab_isg_eq_4}, for $d = 2$, $3$, and $4$,
respectively. For all runs we ensure that the parallel tempering Monte Carlo
moves have acceptance probabilities of at least 30\%.

\begin{table}
\caption{
Parameters of the equilibrium simulations for the two-dimensional gauge glass.
$N_{\mathrm{samp}}$ is the number of samples, $N_{\mathrm{sweep}}$ is the 
total number of Monte Carlo sweeps for each of the $2 N_T$ copies (two 
replicas per temperature) for a single sample, and $N_T$ is the number of 
temperatures used in the parallel tempering method. The lowest temperature 
used is $0.13$, the highest $1.058$. For $L = 24$ the lowest temperature 
studied is $0.20$.
\label{tab_gg_eq_2}
}
\begin{tabular*}{\columnwidth}{@{\extracolsep{\fill}} c r c l }
\hline
\hline
$L$  &  $N_{\mathrm{samp}}$  & $N_{\mathrm{sweep}}$ & $N_T$  \\
\hline
3  & $10\; 000$ & $8.0 \times 10^4$ & 30 \\
4  & $10\; 400$ & $8.0 \times 10^4$ & 30 \\
6  & $10\; 150$ & $8.0 \times 10^4$ & 30 \\
8  &  $8\; 495$ & $2.0 \times 10^5$ & 30 \\
12 &  $6\; 890$ & $8.0 \times 10^5$ & 30 \\
16 &  $2\; 500$ & $2.0 \times 10^6$ & 30 \\
24 &  $2\; 166$ & $2.0 \times 10^6$ & 24 \\
\hline
\hline
\end{tabular*}
\end{table}

\begin{table}
\caption{
Parameters of the equilibrium simulations for the three-dimensional gauge
glass. The lowest temperature simulated is $0.05$, the highest $0.947$. 
The different quantities are explained in the caption of 
Table \ref{tab_gg_eq_2}.
\label{tab_gg_eq_3}
}
\begin{tabular*}{\columnwidth}{@{\extracolsep{\fill}} c r c l }
\hline
\hline
$L$  &  $N_{\mathrm{samp}}$  & $N_{\mathrm{sweep}}$ & $N_T$  \\
\hline
3 & $10\; 000$ & $6.0 \times 10^3$ & 53 \\
4 & $10\; 000$ & $2.0 \times 10^4$ & 53 \\
5 & $10\; 000$ & $6.0 \times 10^4$ & 53 \\
6 & $ 5000$ & $2.0 \times 10^5$ & 53 \\
8 & $ 2000$ & $1.2 \times 10^6$ & 53 \\
\hline
\hline
\end{tabular*}
\end{table}

\begin{table}
\caption{
Parameters of the equilibrium simulations for the gauge glass in four 
dimensions. The lowest temperature used is $0.70$, the highest $1.345$.
The different quantities are explained in the caption of 
Table \ref{tab_gg_eq_2}.
\label{tab_gg_eq_4}
}
\begin{tabular*}{\columnwidth}{@{\extracolsep{\fill}} c r c l }
\hline
\hline
$L$  &  $N_{\mathrm{samp}}$  & $N_{\mathrm{sweep}}$ & $N_T$  \\
\hline
3 & $5000$ & $2.0 \times 10^4$ & 17 \\
4 & $5000$ & $8.0 \times 10^4$ & 17 \\
5 & $5000$ & $4.0 \times 10^5$ & 17 \\
\hline
\hline
\end{tabular*}
\end{table}

\begin{table}
\caption{
Parameters of the equilibrium simulations for the Ising spin glass in two
dimensions. The different quantities are explained in the caption of 
Table \ref{tab_gg_eq_2}. For $L = 128$ the minimum temperature used is $0.20$.
For all other system sizes the minimum temperature is $0.05$. The maximum
temperature used is $1.391$.
Note that for $L \le 16$ standard parallel tempering Monte Carlo is used, 
whereas for $L \ge 32$ the cluster method by Houdayer 
(Ref.~\onlinecite{houdayer:01}) is applied.
\label{tab_isg_eq_2}
}
\begin{tabular*}{\columnwidth}{@{\extracolsep{\fill}} c r r l }
\hline
\hline
$L$  &  $N_{\mathrm{samp}}$  & $N_{\mathrm{sweep}}$ & $N_{T}$
\\
\hline
  3 & $10\; 000$ & $2.0 \times 10^5$ & 20 \\
  4 & $10\; 000$ & $2.0 \times 10^5$ & 20 \\
  6 & $10\; 000$ & $2.0 \times 10^5$ & 20 \\
  8 & $10\; 000$ & $2.0 \times 10^5$ & 20 \\
 10 & $10\; 000$ & $1.0 \times 10^6$ & 20 \\
 12 & $10\; 000$ & $1.0 \times 10^6$ & 20 \\
 16 & $10\; 000$ & $1.0 \times 10^6$ & 20 \\
 24 & $10\; 000$ & $1.0 \times 10^5$ & 20 \\
 32 & $10\; 000$ & $1.0 \times 10^5$ & 20 \\
 64 &  $1000$ & $1.0 \times 10^6$ & 40 \\
128 &      $250$ & $1.0 \times 10^6$ & 63 \\
\hline
\hline
\end{tabular*}
\end{table}

\begin{table}
\caption{
Parameters of the equilibrium simulations for the Ising spin glass in three
dimensions. The lowest temperature used in the parallel tempering method is
$0.10$, the highest $2.00$. Additional higher temperatures have been computed
with simple Monte Carlo.
The different quantities are explained in the caption of 
Table \ref{tab_gg_eq_2}.
\label{tab_isg_eq_3}
}
\begin{tabular*}{\columnwidth}{@{\extracolsep{\fill}} c r r l }
\hline
\hline
$L$  &  $N_{\mathrm{samp}}$  & $N_{\mathrm{sweep}}$ & $N_T$  \\
\hline
3  & $20\; 000$ & $5.0 \times 10^3$ & 18 \\
4  & $30\; 000$ & $5.0 \times 10^3$ & 18 \\
5  & $10\; 000$ & $5.0 \times 10^4$ & 18 \\
6  & $10\; 000$ & $1.5 \times 10^5$ & 18 \\
8  & $10\; 000$ & $5.0 \times 10^6$ & 18 \\
\hline
\hline
\end{tabular*}
\end{table}

\begin{table}
\caption{
Parameters of the equilibrium simulations for the Ising spin glass in four
dimensions. The lowest temperature used in the parallel tempering method is 
$0.20$, the highest $2.80$. Additional higher temperatures have been computed
with simple Monte Carlo. 
The different quantities are explained in the caption of 
Table \ref{tab_gg_eq_2}.
\label{tab_isg_eq_4}
}
\begin{tabular*}{\columnwidth}{@{\extracolsep{\fill}} c r c l }
\hline
\hline
$L$  &  $N_{\mathrm{samp}}$  & $N_{\mathrm{sweep}}$ & $N_T$  \\
\hline
3  & $60\; 000$ & $6.0 \times 10^3$ & 12 \\
4  & $30\; 000$ & $6.0 \times 10^4$ & 12 \\
5  & $12\; 190$ & $3.0 \times 10^5$ & 23 \\
\hline
\hline
\end{tabular*}
\end{table}

Parameters used in the nonequilibrium simulations are summarized in Tables
\ref{tab_gg_neq} and \ref{tab_isg_neq} for the GG, as well as the ISG, 
respectively.

\begin{table}
\caption{
Parameters of the off-equilibrium simulations as a function of space
dimension $d$ for the GG. $L_{\rm b}$ is the size of the system used 
and $t_{\rm w}$ is the ``waiting time.''  $N_{\mathrm{samp}}$ is the number
of samples used for the disorder average. The same temperature sets as in the
equilibrium simulations have been used.
\label{tab_gg_neq}
}
\begin{tabular*}{\columnwidth}{@{\extracolsep{\fill}} c r c c }
\hline
\hline
$d$ & $L_{\rm b}$ & $N_{\mathrm{samp}}$ & $t_{\rm w}$ \\
\hline
$2$ & $64$ & $1000$ & $1.638 \times 10^4$ \\
$3$ & $16$ & $1000$ & $8.192 \times 10^3$ \\
$4$ & $10$ & $1000$ & $1.310 \times 10^5$ \\
\hline
\hline
\end{tabular*}
\end{table}

\begin{table}
\caption{
Parameters of the off-equilibrium simulations as a function of space
dimension $d$ for the ISG. The different quantities are explained in
the caption of Table \ref{tab_gg_neq}. The same temperature sets as in the
equilibrium simulations have been used.
\label{tab_isg_neq}
}
\begin{tabular*}{\columnwidth}{@{\extracolsep{\fill}} c r c c }
\hline
\hline
$d$ & $L_{\rm b}$ & $N_{\mathrm{samp}}$ & $t_{\rm w}$ \\
\hline
$2$ & $128$ & $1000$ & $32\; 768$ \\
$3$ &  $24$ & $1000$ & $32\; 768$ \\
$4$ &  $10$ & $1000$ & $32\; 768$ \\
\hline
\hline
\end{tabular*}
\end{table}

By convention, values quoted at time $t$
correspond to an average taken between MCS's $[t+1]$ to $[2t]$ following an
anneal of $t$ MCS's.

\bibliography{refs,comment}

\begin{thebibliography}{63}
\bibitem[{com({\natexlab{a}})}]{comment:convention}
\bibinfo{note}{{Note that we use the convention by Berthier and Young
  (Ref.~\onlinecite{berthier:04}) where $t_{\rm w}$ represents a time after
  quench, whereas $t$ refers to a time difference}}.
\bibitem[{com({\natexlab{b}})}]{comment:figs}
\bibinfo{note}{{In order to save space only representative figures for the
  two-dimensional gauge glass are displayed}}.
\bibitem[{com({\natexlab{c}})}]{comment:4d}
\bibinfo{note}{{A similar analysis in four dimensions (not shown) shows
  qualitatively similar results}}.
\end{thebibliography}

\end{document}